\documentclass[pra,aps,showpacs]{revtex4}
\usepackage{graphicx}
\begin{document}
\newcommand{\be}{\begin{equation}}
\newcommand{\nn}{\nonumber}
\newcommand{\ee}{\end{equation}}
\newcommand{\bea}{\begin{eqnarray}}
\newcommand{\eea}{\end{eqnarray}}
\newcommand{\wee}[2]{\mbox{$\frac{#1}{#2}$}}   
\newcommand{\unit}[1]{\,\mbox{#1}}
\newcommand{\degree}{\mbox{$^{\circ}$}}
\newcommand{\ltish}{\raisebox{-0.4ex}{$\,\stackrel{<}{\scriptstyle\sim}$}}
\newcommand{\bin}[2]{\left(\begin{array}{c} #1 \\ #2\end{array}\right)}
\newcommand{\p}{_{\mbox{\small{p}}}}
\newcommand{\m}{_{\mbox{\small{m}}}}
\newcommand{\tra}{\mbox{Tr}}
\newcommand{\rs}[1]{_{\mbox{\tiny{#1}}}}        
\newcommand{\ru}[1]{^{\mbox{\small{#1}}}}

\title{Retrodictive fidelities for pure state postselectors} 
\author{John Jeffers}
\affiliation{SUPA, Department of Physics, University of Strathclyde, John Anderson Building, 107 Rottenrow, Glasgow G4 0NG, UK. john@phys.strath.ac.uk}

\begin{abstract}
Two measures of fidelity are proposed for postselecting devices, the retrodictive conditional probability that the state in the measurement arm is the one indicated by the detectors, and the probability that the device produces the state that it would produce if working perfectly. The first is the natural quantity that one wishes to maximise to improve the device operation. The second corresponds more closely with the accurate operation of the device than the more usual overlap-based fidelity. The results are particularly applicable to the types of state preparation and logic gate operation in linear optical quantum computing.
\end{abstract}


\maketitle

\section{Introduction}
Fidelity is one of the fundamental concepts in quantum information. It applies to any device which manufactures states of a quantum system, and is unable to do so perfectly. The fidelity should quantify the ability of the device to produce the desired state. Various definitions abound, but if the desired state is pure then the one which is often used is
\bea
\label{overlap}
F_o=\tra[\hat{\rho}\hat{\rho}^\prime] = \langle \psi |\hat{\rho}^\prime|\psi\rangle,
\eea
where $\hat{\rho}=|\psi\rangle \langle \psi |$ is the desired state and $\hat{\rho}^\prime$ is the state which is actually made \cite{Jozsa94}. This measure has the appealingly simple interpretation that it is the probability that the state $\hat{\rho}^\prime$ gives the result $|\psi\rangle$ when measured by a detector which includes this state as one of its possible results. If both states are pure the fidelity is the square of the overlap of the two states, and so here this quantity will be called the overlap fidelity. Mixed state fidelities are more complicated, but Jozsa has proposed a measure based on Uhlmann's transition probability which has the required properties of a fidelity, and reduces to the overlap fidelity when one of the states is pure \cite{Jozsa94,Uhlmann76}. Others use the square root of this quantity as a measure of the distance between (or perhaps, the closeness of) the two states in Hilbert space \cite{Uhlmann76,Fuchs96,Nielsen00}.

Postselecting devices produce particular states based on the results of measurements of some of their outputs. They are especially useful in optics, where it is often the case that prior to measurement the state of a system is mixed. In particular it has been suggested that optical postselecting devices could be cascaded to make the gates necessary for a scalable linear optical quantum computer (LOQC) \cite{KLM01,Kok1}. Other possible schemes such as the one-way quantum computer rely on the results of measurements on cluster states in optics \cite{Nielsen, Browne}, and in matter-based schemes \cite{Joo,Louis,Kok2}. For any quantum information-based device to work the postselectors must have very high fidelities, and this is especially true of LOQC due to the cascading of devices required. Unfortunately, photodetectors are relatively poor, and so they drastically limit the quality of state production. Even a modest optical quantum information processor is therefore beyond the reach of present day technology. Imperfections in other components also reduce the fidelity, but detector imperfection is the main limitation. 

The important criterion for any machine is whether or not it performs the required task, so a reasonable criterion for a postselecting device is therefore whether or not its output would perform the task for which it was designed. For some postselecting devices the required output is `sacred': only that part of the output which corresponds to perfect device operation will perform the relevant quantum information task. Other devices are less sensitive, and can partially tolerate some of the possible incorrect output states. 

In this paper the overlap fidelity is calculated for a postselecting device subject to imperfect detection. The calculation shows directly the relation between this quantity and the retrodictive conditional probabilities (RCPs) that particular states were present in the detected modes, given particular detection results at the detectors. The overlap fidelity is larger than the probability that the device produces the desired state, so the latter quantity, which is dubbed here the correct output fidelity, is proposed as a measure of fidelity. This definition corresponds closely with the device output criterion described in the previous paragraph. A useful, simply-calculated lower bound on the correct output fidelity, the RCP that the detection arm contains the state that the detector indicates, will often correspond precisely with the correct output fidelity. 

The paper is organised as follows. Section 2 analyses the perfect postselecting device, which is assumed to make a pure state when the detector result indicates that the device has functioned correctly. In Section 3 the effect of imperfect detection on the output state made by the device is calculated. This imperfect state is used to calculate the overlap fidelity, and the correct output fidelity. It is shown that both can be written usefully in terms of a sum of products of retrodictive conditional probabilities and overlaps. For the most important of these terms the overlap is perfect and the fidelity reduces to a simple probability.
Then the results are illustrated using a postselecting device which could form an optical gate in a LOQC. In the final section the results are summarised and conclusions are provided. 

\section{Perfect postselector}
\begin{figure}[htbp]
\centering
\includegraphics[totalheight=3cm]{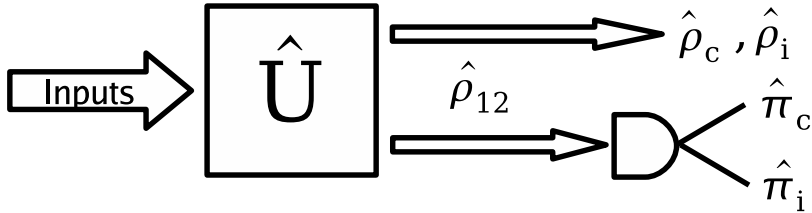}
\caption{A perfect postselecting device}
\label{fig1}
\end{figure}

Consider the general postselecting device pictured schematically in Fig. \ref{fig1}. A multimode input state is fed into a box which performs perfectly a known unitary transformation $\hat{U}$. The output of the box is divided into two parts. One part of the output (mode 1) will form the postselecting device output. The other part (mode 2) will be measured. Either mode 1, or mode 2, (or both) may be compound modes. The output state of the unitary transforming box is $\hat{\rho}_{12}$. The particular input states of the device, together with the exact form of the unitary transformation, determine this state, but they do not matter otherwise, and will take no further part in the general theory. 

Mode 2 is measured by a perfect detector (or set of perfect detectors for a compound mode); in other words the measured state is the same state as that indicated by the detector. The measurement in mode 2 has only two possible results, the required correct result $c$ and the incorrect result $i$. These are described by a probability operator measure (POM) with corresponding elements $\hat{\pi}_c$ and $\hat{\pi}_i$ which satisfy
\bea
\hat{\pi}_c + \hat{\pi}_i = \hat{1},
\eea
the identity operator for the system. This criterion is the mathematical representation of the fact that the detector must produce a result. Normally there will be many possible incorrect results, and these can either be lumped together to form a single incorrect POM element or treated separately. One simplification made at this stage is that the detector results are mutually orthogonal. This will often be the case for practical devices, particularly in optics, where the photon number states typically form the ideal measurement basis.

When the detector fires with result $c$ the state $\hat{\rho}_c$ is produced in mode 1, and when the result $i$ is obtained the state $\hat{\rho}_i$ is produced. These states are simply found as
\bea
\label{rhotick}
\hat{\rho}_c &=& \frac {\tra_2 \left[ \hat{\rho}_{12} \hat{\pi}_c \right]}{\tra_{12} \left[ \hat{\rho}_{12} \hat{\pi}_c \right]} = \frac {\tra_2 \left[ \hat{\rho}_{12} \hat{\pi}_c \right]}{\tra_2 \left[ \hat{\rho}_2 \hat{\pi}_c \right]}\\
\label{rhocross}
\hat{\rho}_i &=& \frac {\tra_2 \left[ \hat{\rho}_{12} \hat{\pi}_i \right]}{\tra_{12} \left[ \hat{\rho}_{12} \hat{\pi}_i \right]} = \frac {\tra_2 \left[ \hat{\rho}_{12} \hat{\pi}_i \right]}{\tra_2 \left[ \hat{\rho}_2 \hat{\pi}_i \right]},
\eea
where $\hat{\rho}_2=\tra_1 \hat{\rho}_{12}$ is the output state in mode 2 alone without any knowledge of the state in mode 1. The denominators of these equations are the probabilities of obtaining the results $c$ and $i$. The measurement results in mode 2 provide an effective basis for this arm (orthogonal if the measurement results are orthogonal). The probabilities of obtaining these results are the important quantities, and so nothing is lost if the state $\hat{\Lambda}_2$ is substituted for $\hat{\rho}_2$ in eqs. (\ref{rhotick}) and (\ref{rhocross}), where
\bea
\label{lambda}
\hat{\Lambda}_2 = p_c \frac{\hat{\pi}_c}{\tra \hat{\pi}_c} + p_i \frac{\hat{\pi}_i}{\tra \hat{\pi}_i}= p_c \hat{\Lambda}_c +p_i \hat{\Lambda}_i.
\eea
$p_c$ and $p_i$ are then the denominators in eqns. (\ref{rhotick}) and (\ref{rhocross}) and so form the probabilities of obtaining these results. In a loose sense, they are the `{\it a priori}' probabilities that the device produced the states $\hat{\Lambda}_c$ and $\hat{\Lambda}_i$ in mode 2 \footnote{These are {\it not} strictly the {\it a priori} probabilities that the device produced these particular states, as the device really produces an entangled state of the two modes. It can never produce the unentangled states of mode 2, $\hat{\Lambda}_c$ and $\hat{\Lambda}_i$, so such a decomposition is fictional. For a full discussion of this point see \cite{Pegg05}.}. This fictional decomposition will be used in Section 3. It is possible, however, to give a meaning to the operators $\hat{\pi}_c/\tra (\hat{\pi}_c)$ and $\hat{\pi}_i/\tra (\hat{\pi}_i)$ within retrodictive quantum theory, where they form the retrodictive states, the best descriptions of the pre-measurement system in mode 2 given knowledge \emph{only} of the measurement outcome \cite{Pegg02a}. 

For the output of the device to be most useful for quantum information processing it will almost always be the case that under perfect operation a pure state will be required in mode 1. If the state $\hat{\rho}_c$ is to be pure then the measurement result $\hat{\pi}_c$ must correspond to a pure state of mode 2. No such restriction is imposed on the incorrect states or POM elements.

\section{Imperfect postselector}
\subsection{Output state}
\begin{figure}[htbp]
\centering
\includegraphics[totalheight=4cm]{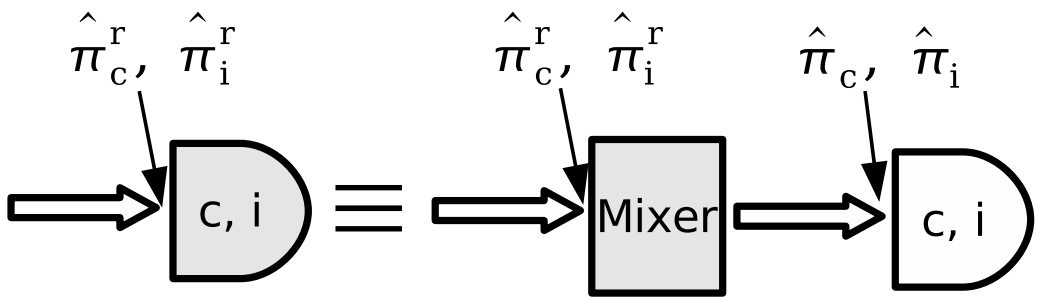}
\caption{An imperfect measurement device}
\label{fig2}
\end{figure}

Imperfect postselection is largely a consequence of imperfect measurement, where the measured state is not precisely that which is indicated by the detector result. Here an imperfect detector is modelled by a perfect detector preceded by a state mixer (Fig. \ref{fig2}). One example of this is the modelling of an imperfect photodetector of a particular quantum efficiency by a perfect photodetector preceded by an attenuator whose transmission factor is the quantum efficiency \cite{Yuen80}. 
Imperfect photodetection has been shown to be equivalent to projection onto mixed states of the electromagnetic field \cite{Barnett98}. 

There is a choice in the method of calculation. Either the evolution of the state $\hat{\Lambda}_2$ forwards through the mixer, and hence the probabilities of particular measurement results at the perfect detector, can be calculated, or the perfect detector POM elements can be evolved backwards through the mixer, to find mixed POM elements. These mixed POM elements are effectively the measurement results for the imperfect detector \cite{Barnett98}. The latter option is chosen here partially because of this simple interpretation, but also because for real systems the calculations will be slightly simpler due to the trivial nature of the POM elements of the perfect detector. 

The backwards evolution is performed using retrodictive quantum theory \cite{Pegg02b}, and the calculation will depend on the particular form of detector and mixer, but the form of the mixed POM elements, for orthogonal $\hat{\pi}_c$ and $\hat{\pi}_i$, will be \footnote{In real systems such as typically encountered in optics the evolution will always be describable by a type of Lindblad evolution.}
\bea
\label{pitick1}
\hat{\pi}^r_c = \pi^r(c|c)\hat{\pi}_c + \pi^r(i|c)\hat{\pi}_i,\\
\label{picross1}
\hat{\pi}^r_i = \pi^r(c|i)\hat{\pi}_c + \pi^r(i|i)\hat{\pi}_i,
\eea
where, for example, $\pi^r(i|c)$ is the retrodictive `weight' that the system is projected onto the state corresponding to the POM element  $\hat{\pi}_i$ given that the detector result was $c$. Again, there is always a measurement result, and this leads to the conditions
\bea
\pi^r(c|c)+\pi^r(c|i) = \pi^r(i|c) + \pi^r(i|i) = 1.
\eea
These conditions may be unfamiliar, but in the Appendix it is shown that they arise because for any mixing device the retrodictive weights are equivalent to corresponding predictive probabilities, e.g.
\bea
\pi^r(i|c) \equiv P^p(c|i),
\eea
where $P^p(c|i)$ is the predictive probability that the state $\hat{\Lambda}_i$ evolves forwards in time, under the action of the mixer, into the state $\hat{\Lambda}_c$. The equivalence is not correct for multiple nonorthogonal measurement results, but then it is possble to use the retrodictive master equation \cite{Barnett01,Pegg02b} to calculate the evolution of the POM elements. Thus eqns. (\ref{pitick1}) and (\ref{picross1}) can be rewritten in terms of the predictive probabilities as
\bea
\label{pitick2}
\hat{\pi}^r_c = P^p(c|c)\hat{\pi}_c + P^p(c|i)\hat{\pi}_i,\\
\label{picross2}
\hat{\pi}^r_i = P^p(i|c)\hat{\pi}_c + P^p(i|i)\hat{\pi}_i.
\eea

The state produced by the device when the detector fires with the correct result, $\hat{\rho}_c^\prime$ is calculated simply, by substituting eq. (\ref{pitick2}) for $\hat{\pi}_c$ in eq. (\ref{rhotick}) to find
\bea
\label{rhoprimed}
\nn \hat{\rho}_c^\prime &=& \frac
{\tra_2 \left[ \hat{\rho}_{12} \hat{\pi}^r_c \right]}
{\tra_2 \left[ \hat{\Lambda}_2 \hat{\pi}^r_c \right]} = \frac
{P^p(c|c)\tra_2 \left[ \hat{\rho}_{12} \hat{\pi}_c \right]
+ P^p(c|i) \tra_2 \left[ \hat{\rho}_{12} \hat{\pi}_i \right]}
{ p_c P^p(c|c)\tra_2 \left[ \hat{\Lambda}_c \hat{\pi}_c \right]
+ p_i P^p(c|i) \tra_2 \left[ \hat{\Lambda}_i \hat{\pi}_i \right]}\\
&=& \frac
{P^p(c|c)\tra_2 \left[ \hat{\rho}_{12} \hat{\pi}_c \right]
+ P^p(c|i) \tra_2 \left[ \hat{\rho}_{12} \hat{\pi}_i \right]}
{ p_c P^p(c|c)+ p_i P^p(c|i)}.
\eea
The denominator can be recognised as the probability that the imperfect detector fires with the outcome $c$. A similar result to eq. (\ref{rhoprimed}) can be found for the incorrect detector outcome $i$. Of course it is hoped that this imperfect state corresponds as closely as possible to the state produced by the perfect postselector described in the previous section. The closeness of this correspondence is quantified by the fidelity, which is calculated below. 

One final point for this subsection is that if there are $N$ separate incorrect detector results it is possible to write eq. (\ref{rhoprimed}) as
\bea
\hat{\rho}_c^\prime = \frac
{P^p(c|c)\tra_2 \left[ \hat{\rho}_{12} \hat{\pi}_c \right]
+ \sum_{j=1}^N P^p(c|i,j) \tra_2 \left[ \hat{\rho}_{12} \hat{\pi}_{i,j} \right]}
{ p_c P^p(c|c)+ \sum_{j=1}^N p_{i,j} P^p(c|i,j)}. 
\eea
Thus it is always possible to write the imperfect state produced by the device in terms of the states produced by the perfect device and their corresponding predictive conditional probabilities.

\subsection{Fidelity}
The overlap between the pure state produced by the perfect device and the state produced by the imperfect device is normally used as a measure of fidelity. For the states derived above this is found, using eqns. (\ref{rhotick}) and (\ref{lambda}) and the property that the trace of the square of a pure state is unity, to be
\bea
\nn F_o = \tra_1[\hat{\rho}_c \hat{\rho}_c^\prime]
&=& \frac
{P^p(c|c) \tra_1\left\{ \tra_2 \left[ \hat{\rho}_{12} \hat{\pi}_c \right] \right\}^2/p_c}
{ p_c P^p(c|c)+ p_i P^p(c|i)} +\frac
{P^p(c|i) \tra_1\left\{ \tra_2 \left[ \hat{\rho}_{12} \hat{\pi}_i \right]\tra_2 \left[ \hat{\rho}_{12} \hat{\pi}_i \right]\right\}/p_c}
{ p_c P^p(c|c)+ p_i P^p(c|i)}\\
\nn &=& \frac
{p_c P^p(c|c) + p_i P^p(c|i) \tra_1[\hat{\rho}_c\hat{\rho}_i]}
{ p_c P^p(c|c)+ p_i P^p(c|i)}\\
&=& P^{retr}(c|c) + P^{retr}(i|c)\tra_1[\hat{\rho}_c\hat{\rho}_i],
\eea
where Bayes' Theorem \cite{Bayes} has been used, and where e.g. $P^{retr}(c|c)$ is the RCP that state $\hat{\Lambda}_c$ was ``present" in mode 2 given that the detector result was $c$, a quantity which takes account of the {\it a priori} probability of such a state being present in mode 2. 

The first term in the overlap fidelity as calculated above is the probability that the state in the detector mode is the same as the one indicated by the imperfect detector. It does not depend formally on the state which the device produces in mode 1, only on the {\it a priori} probabilities of the two mode 2 states in the detector-basis decomposition and the properties of the mixing device. As the detector improves, this probability approaches unity. This term is here dubbed the {\it retrodictive fidelity}, $F_r=P^{retr}(c|c)$. 

The second term in the overlap fidelity is proportional to the probability that the state in the detector mode is not the same as the one indicated by the imperfect detector. Provided that the two output states in mode 1 $\hat{\rho}_c$ and $\hat{\rho}_i$ are not orthogonal this term makes a contribution to the overlap fidelity, a contribution which of course vanishes for a perfect detector. 

When this second term does not vanish it ought to contribute usefully to the device output, but it will not always do so. Often the postselecting device will only work when the state $\hat{\rho}_c$ is produced. In order to find out how likely this is the state $\hat{\rho}_i$ is decomposed into two parts:
\bea
\label{decomp}
\hat{\rho}_i = P^{max}_i \hat{\rho}_c + \hat{\gamma}_{\bar{c}},
\eea
where $P^{max}_i$ is the maximum amount of $\hat{\rho}_c$ that can be extracted from $\hat{\rho}_i$ \footnote{As a simple example of the meaning of this consider the states $|1\rangle$ and $|+\rangle=\frac{1}{\sqrt{2}}(|1\rangle+|2\rangle)$, where $|1\rangle$ and $|2\rangle$ are orthogonal. The maximum amount of $|1\rangle\langle1|$ that can be extracted from $|+\rangle\langle+|$ is 1/2, but none of $|+\rangle\langle+|$ can be extracted from $|1\rangle\langle1|$, as $|1\rangle$ does not contain any of $|2\rangle$.}, loosely the probability that $\hat{\rho}_i$ {\it is} $\hat{\rho}_c$, and where $\hat{\gamma}_{\bar{c}}$ is the remainder. It is evident that the first term will always make a meaningful contribution to the device output, and that the second will not necessarily do so. Accordingly, the second term in eq. 
(\ref{decomp}) is discarded from the trace in the numerator in the fidelity, and a different measure called here the {\it correct output fidelity}, $F_c$, is defined
\bea
\label{fidelity}
F_c = F_r + P^{retr}(i|c)P^{max}_i.
\eea
This is smaller than the overlap fidelity, but it is a safer quantity to use given that $F_o$ is based on an output state, parts of which are not useful.  The result for multiple incorrect detector results is obvious; from eq. (\ref{fidelity}),  
and the result is 
\bea
F_c = F_r + \sum_{j=1}^N P^{retr}(i,j|c)P^{max}_{i,j}.
\eea

For some devices it will not be possible to perform the decomposition in eq. (\ref{decomp}) for nonzero $P^{max}_i$, and so $F_c$ will then be equal to the retrodictive fidelity $F_r$. Otherwise $F_r$ forms a lower bound on $F_c$. More importantly, it also forms the quantity which should be maximised in order to maximise the overall fidelity, whichever measure is used. Recently a measurement has been introduced to distinguish between linearly-dependent states based on maximising a retrodictive conditional probability of this type, and such measurements have beeen dubbed maximum confidence measurements \cite{Croke1,Croke2}. The relation between the different measures of fidelity is illustrated in the inequality
\bea
F_r \leq F_c \leq F_o \leq 1.
\eea
All of the measures of fidelity are equal if $\hat{\rho}_c \perp \hat{\rho}_i$. 

The correct output fidelity is an accurate measure when a particular device output, and only this output, is required, but for some devices this will not be the case. Then, although the remainder in eq. (\ref{decomp}) cannot be used to make any of the correct state, when the postselecting device is combined with others to make a larger device it is possible for some part of the remainder to contribute to the correct output fidelity of this larger device. This contribution will usually be small, but the ``true" fidelity of the component system ought to contain the effect of such terms. 
\subsection{Example}
\begin{figure}[htbp]
\centering
\includegraphics[totalheight=4.5cm]{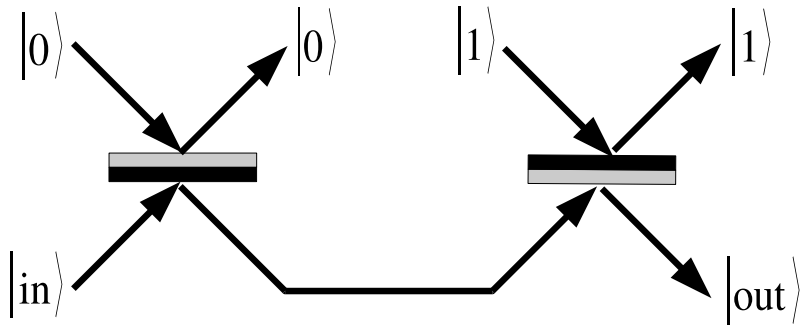}
\caption{Nonlinear sign-shift gate from \cite{Ralph01}. The beam splitter reflectivities are $5-3\sqrt{2}$ (left) and $(3-\sqrt{2})/7$ and there is a phase change of $\pi$ on reflection from the grey side. }
\label{fig3}
\end{figure}

As an example the formalism is applied to the optical nonlinear sign shift gate (NS) introduced by Ralph et al.\cite{Ralph01} and shown in Fig. \ref{fig3}. When working perfectly, this device performs the operation
\bea
\alpha|0\rangle + \beta |1\rangle + \gamma |2\rangle \rightarrow
\alpha|0\rangle + \beta |1\rangle - \gamma |2\rangle.
\eea
\begin{figure}[htbp]
\centering
\includegraphics[totalheight=4.5cm]{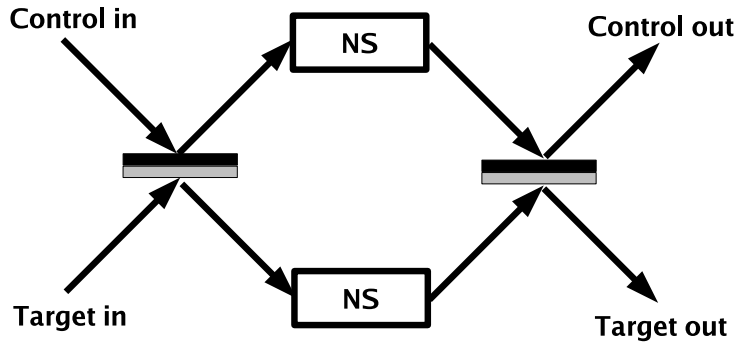}
\caption{Control sign-shift gate from \cite{Lund03}. The beam splitters are both 50/50 and there is a phase change of $\pi$ on reflection from the grey side. }
\label{fig4}
\end{figure}
Two of these gates form the basis of the control sign shift operation (CS), shown in Fig. \ref{fig4}, which could be used as a basic gate for a LOQC \cite{Lund03}. The truth table for the CS gate is as follows: $|00\rangle \rightarrow |00\rangle, |01\rangle \rightarrow |01\rangle, |10\rangle \rightarrow |10\rangle, |11\rangle \rightarrow -|11\rangle$, where the first number in each state is the control photon number and the second is the target photon number. 

\begin{figure}[htbp]
\centering
\includegraphics[totalheight=7cm]{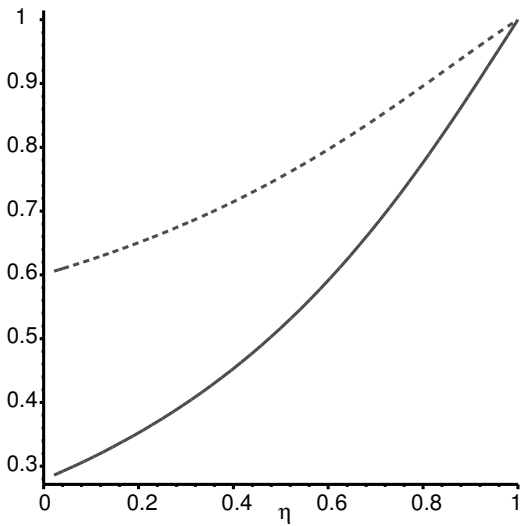}
\caption{Plot of overlap fidelity (dashed) and retrodictive fidelity (solid) against quantum efficiency $\eta$. }
\label{fig5}
\end{figure}

The single measurement output mode in the theory described in previous sections is a double compound mode for the NS gate. The device works when no photons are detected in one of the detector modes, and one is found in the other mode. 
This device is a very good illustration of the overestimate provided by the overlap fidelity, as none of the states produced by the perfect device is orthogonal to the required state, but none of them can be written in terms of it. This means that the correct output fidelity is equivalent to the retrodictive fidelity $F_r$, in this case the probability of the presence of zero and one photon in the appropriate modes given that zero and one photon were indicated by the detectors in these modes. For the sake of simplicity it is assumed that $\alpha=\beta=\gamma=1/\sqrt{3}$, and the details of the calculation, which is straightforward but tedious for the overlap fidelity are omitted. The general forms of the detector probabilities are already known \cite{Jed04, Jed00}. Figure \ref{fig5} shows a plot of the overlap and retrodictive fidelities as a function of detector quantum efficiency. 
It is clear that the retrodictive fidelity, which here gives the probability that the device actually produces the required state, is significantly lower than the overlap fidelity over the whole range of quantum efficiency, apart from at 1, where of course all measures coincide. This clearly demonstrates the peril of using overlap fidelity as a measure for such a device. 

Straightforward analysis and probability theory shows that the fidelity of the CS gate with simple lossy detectors with no dark counts requires both NS gates to function correctly. Otherwise gate errors are introduced \footnote{Outputs which correspond to the detection of no photons in the detector modes in which one photon should be detected will not always introduce a gate error, but such a result means that the NS gate has not worked, so the output will be discarded.  For a detector with a finite dark count rate, such outputs might contribute to the fidelity, but any contribution will be tiny.}. In this case the retrodictive fidelity of the CS will be simply the square of the retrodictive fidelity of the NS. 

\section{Conclusions}

In this paper two measures of fidelity for postselecting devices have been proposed. Each measure is based on the correct functioning of the device. The correct output fidelity $F_c$ can be loosely described as the probability that the device produces the pure state in its output mode that it would produce if it were working perfectly. More correctly it is the maximum amount or fraction of the desired state that one can make out of the imperfect state that is actually produced. This is always lower than the more normally used overlap fidelity. Quantum gates can require particular output states to function correctly, and $F_c$ will normally form the probability that the gate performs the task for which it was designed, given the required detector result. Scalability of gates is safe when based on such a probability, and unsafe when based on an upper bound such as the ovelap fidelity. 

The largest term contributing to $F_c$ is the retrodictive fidelity $F_r$, the retrodictive conditional probability that the state indicated by the detector in the detected mode is the same as the state in the detected mode. In other words it depends on the correct functioning of the detector arm only. This retrodictive fidelity is simple to calculate, does not depend formally on the output of the device, and provides a useful lower bound on $F_c$, with which it will correspond precisely in many cases. In future work simple methods of increasing this probability will be investigated, and the application of this concept of fidelity in the case where the internal components of the device are imperfect will be examined. This latter case can correspond to the postselecting device being formed by performing a nonunitary transformation on the inputs, and will be important for general scaling of the fidelity in composite devices. 

\section*{Acknowledgments}
This work was partially supported by the UK Engineering and Physical Sciences Research Council. I thank Craig Hamilton and Steve Barnett for useful discussions.

\section*{Appendix}
Suppose that a quantum system is prepared, evolves and then is measured. The measurement POM elements $\hat{\pi}_l$ are mutually orthogonal, and each one corresponds precisely to just one of the possible prepared states $\hat{\rho}_k$. The probability that measurement result $l$ is obtained given that the preparation event was $k$ is
\bea
\label{pcpp}
P(l|k) = \tra \left[ \hat{\rho}_k^p \hat{\pi}_l \right],
\eea
where $\hat{\rho}_k^p$ is the predictively evolved density operator (evolved forwards in time), given by
\bea
\label{pdo}
\hat{\rho}_m^p = \sum_k P^p(m|k) \hat{\rho}_k,
\eea
in which $P^p(m|k)$ is the predictive probability that $\hat{\rho}_k$ evolves into $\hat{\rho}_m$. Direct substitution of eq. (\ref{pdo}) into eq. (\ref{pcpp}) gives 
\bea
P(l|k) = P^p(l|k).
\eea 

It is also possible to calculate the probability in eq. (\ref{pcpp}) by not evolving the prepared state forwards, but instead evolving the POM element backwards in time. This amounts to formally changing the time of the measurement to prior to the evolution.
\bea
\label{pcpr}
P(l|k) = \tra \left[ \hat{\rho}_k \hat{\pi}_l^r \right],
\eea
where $\hat{\pi}_l^r$ is the retrodictively evolved POM element, given by
\bea
\label{rpe}
\hat{\pi}_l^r = \sum_m \pi^r(m|l) \hat{\pi}_m,
\eea
in which $\pi^r(m|l)$ is the retrodictive weight that measurement result $l$ corresponds to measurement result $m$ prior to the evolution. Again, direct substitution of eq. (\ref{rpe}) into eq. (\ref{pcpr}) leads to 
\bea
P(l|k) = \pi^r(k|l),
\eea 
and so $P^p(l|k) = \pi^r(k|l)$.

\begin{thebibliography}{10}
\bibitem{Jozsa94} R. Jozsa, J. Mod. Opt. {\bf 41}, 2315 (1994).
\bibitem{Uhlmann76} A. Uhlmann, Rep. Math. Phys. {\bf 9}, 273 (1976).
\bibitem{Fuchs96} C.A. Fuchs, Ph.D. Thesis, The University of New Mexico, Albuquerque, New Mexico (1996), quant-ph/9601020.
\bibitem{Nielsen00} M.A. Nielsen and I.L. Chuang, {\it Quantum Computation and Quantum Information} (Cambridge, Cambridge University Press, 2000).
\bibitem{KLM01} E. Knill, R. Laflamme and G. Milburn, Nature {\bf 409}, 46 (2001).
\bibitem{Kok1} For a recent review article with extensive references, see P. Kok {et al.} quant-ph/0512071. 
\bibitem{Nielsen} M.A Nielsen, Phys. Rev. Lett. {\bf 93}, 040503 (2004).
\bibitem{Browne} D.E Browne and T. Rudolph, Phys Rev. Lett. {95}, 010501 (2005).
\bibitem{Joo} J. Joo, Y.L. Lim, A. Beige and P.L. Knight, quant-ph/0601100.
\bibitem{Louis} S.G.R. Louis, K. Nemoto, W.J. Munro and T.P. Spiller, quant-ph/0607060.
\bibitem{Kok2} P. Kok, S.D. Barrett and T.P. Spiller, J. Opt. B: Quantum Semiclass. Opt. {\bf 7}, S166 (2005).
\bibitem{Pegg05}  D.T. Pegg and J. Jeffers, J. Mod. Opt. {\bf 52}, 1835 (2005).
\bibitem{Pegg02a}  D.T. Pegg, S.M. Barnett and J. Jeffers, J. Mod. Opt. {\bf 49}, 913 (2002).
\bibitem{Yuen80} H.P. Yuen and J.H. Shapiro, IEEE Trans. Inf. Theor. {\bf 26}, 78 (1980); R. Loudon, {\it The Quantum Theory of Light}, 3rd Edition (Oxford, Oxford University Press, 2000) p271ff.
\bibitem{Barnett98}  S.M. Barnett, L.S. Phillips and D.T. Pegg, Opt. Commun. {\bf 158}, 45 (1998).
\bibitem{Pegg02b}  D.T. Pegg, S.M. Barnett and J. Jeffers, Phys. Rev. A {\bf 66}, 022106 (2002).
\bibitem{Barnett01} S.M. Barnett, D.T. Pegg, J. Jeffers and O. Jedrkiewicz, Phys. Rev. Lett. {\bf 86}, 2455 (2001). 
\bibitem{Bayes} G.E.P. Box and G.C. Tiao, {\it Bayesian Inference in Statistical Analysis} (Sydney, Addison-Wesley, 1973).
\bibitem{Croke1} S. Croke et al. Phys. Rev. Lett. {\bf 96}, 070401 (2006).
\bibitem{Croke2} P. J. Mosley, S. Croke, I. A. Walmsley and S. M. Barnett, to appear in Phys. Rev. Lett.; quant-ph/0610237.
\bibitem{Ralph01} T.C. Ralph, A.G. White, W.J. Munro and G.J. Milburn, Phys. Rev. A {\bf 65}, 012314 (2001).
\bibitem{Lund03} A.P. Lund, T.B. Bell and T.C. Ralph, Phys.Rev. A {\bf 68}, 022313 (2003). 
\bibitem{Jed04} O. Jedrkiewicz, R. Loudon and J. Jeffers, Phys. Rev. A {\bf 70}, 033805 (2004).
\bibitem{Jed00} S.M. Barnett, D.T. Pegg, J. Jeffers, O. Jedrkiewicz and R. Loudon, Phys. Rev. A {\bf 62}, 022313 (2000). 

\end{thebibliography}
\end{document}